\documentclass[12pt]{article}

\usepackage{graphicx}  
\begin{document}

\begin{center}

{\large\bf Effect of heterogeneity and spatial correlations on the structure of tumor invasion front in cellular 
environments}

\bigskip
Youness Azimzade,$^1$ Abbas Ali Saberi,$^{\dagger,1,2}$ and Muhammad Sahimi$^3$

{\it $^1$Department of Physics, University of Tehran, Tehran 14395-547, Iran}

{\it $^2$Institut f\"ur Theoretische Physik, Universitat zu K\"oln, Z\"ulpicher Strasse 77, 50937 K\"oln, Germany}

{\it $^3$Mork Family Department of Chemical Engineering and Materials Science, University of Southern California, Los 
Angeles, California 90089-1211, USA}

\end{center}

Analysis of invasion front has been widely used to decipher biological properties, as well as the growth dynamics 
of the corresponding populations. Likewise, the invasion front of tumors has been investigated, from which insights 
into the biological mechanisms of tumor growth have been gained. We develop a model to study how tumors' invasion 
front depends on the relevant properties of a cellular environment. To do so, we develop a model based on a nonlinear
reaction-diffusion equation, the Fisher-Kolmogorov-Petrovsky-Piskunov (FKPP) equation, to model tumor growth. Our study 
aims to understand how heterogeneity in the cellular environment's stiffness, as well as spatial correlations in its 
morphology, the existence of both of which has been demonstrated by experiments, affects the properties of tumor 
invasion front. It is demonstrated that three important factors affect the properties of the front, namely, the spatial 
distribution of the local diffusion coefficients, the spatial correlations between them, and the ratio of the cells' 
duplication rate and their average diffusion coefficient. Analyzing the scaling properties of tumor invasion front 
computed by solving the governing equation, we show that, contrary to several previous claims, the invasion front of 
tumors and cancerous cell colonies cannot be described by the well-known models of kinetic growth, such as the 
Kardar-Parisi-Zhang equation.

\newpage

\begin{center}
{\bf I. INTRODUCTION}
\end{center}

Physical properties of tumor microenvironment regulate their functions, such as cell migration through stimulation 
of internal signalings and/or direct manipulations [1-4]. Among the physical features, local rigidity of the tumor 
microenvironment plays an important role in regulating cellular activities [5-8], and more specifically their 
migration, invasion into new cellular media and eventually tumor spreading [9,10]. The effect of the stiffness of 
the tumor microenvironment on cell migration and the properties of the tumnor invasion front has been studied widely
in the literature [11-13], and since rigid environments are difficult to penetrate, proteolytic cellular migrations 
have been reported to be slower in such media [14-17]. 

In all the previous studies, however, the stiffness of the tissues was assumed to be uniform, whereas, according to 
several experimental studies [18-20], it varies spatially in a cellular environment. The fluctuations in the
stiffness appear to be spatially correlated [20,21], but, to our knowledge, neither the type of the correlations 
nor their effect on tumor invasion has been studied [22-24]. THis is despite the fact that the structure of a tumor 
invasion front has been investigated for a rather long time, because it is believed that determining the front's 
properties would shed light on the mechanisms behind its growth [25,26], and would enable one to predict its 
behavior [27,28]. In particular, the surface regularity of brain tumors was recently proposed as a strong predictive
indication of a patient's chance of survival [29,30]. Specifically, a rougher, possibly self-affine, surface of
tumors was associated with shorter survival time. As such, a quantitative understanding of the properties of tumors'
boundaries and their dependence on the relevant physical parameters of the cellular environment may help predicting
its behavior.

The perimeter of a tumor, representing a growing surface, has been studied by dynamic scaling analysis, which is 
the standard approach in the studies of growth of various types of such surfaces, and for which the celebrated 
Kardar-Parisi-Zhang (KPZ) equation [31] has been used. Consider the growth of a surface from a substrate of linear 
size $L$, and assume that each point on the growing surface has a height of $h_i(t)$ with an average height 
$\bar{h}(t)$, while the surface's width is $W(L,t)$, defined by 
\begin{equation}
W^2(L,t)=\frac{1}{N}\sum_i^N(h_i-\bar{h})^2\;.
\end{equation}
For the KPZ-type surface growth, one has
\begin{equation}
W(L,t)\approx L^\alpha f(t/L^z)\;,
\end{equation}
where $f(u)$ is a scaling function such that, $f(u)\propto u^\beta$ if $u\ll 1$, and $f(u)\approx$ constant for
$u\gg 1$, so that for a fixed $L$
\begin{equation}
W(L,t)\propto t^\beta\;.
\end{equation}
$\alpha$ and $\beta$ are, respectively, the surfaqce roughness and growth exponents, and $z=\alpha/\beta$ is the 
dynamic exponent. For the universality class of the KPZ equation in (1+1) dimension one has, $\alpha=1/2$, $\beta=
1/3$, and $z=3/2$. 

Early studies of scaling properties of tumor boundaries revealed their fractal structure with a fractal dimension 
of $D_f\approx 1.21\pm 0.05$, with their scaling properties of the growing surface to be compatible with the 
universality class of molecular beam epitaxy [25] with a roughness exponent $\alpha=0$. The analogy between the two 
phenomena led to the conclusion that a tumor grows linearly and its proliferation is limited to its boundary, with 
marginal cells diffusing on the tumor's front [26]. Later studies indicated that the scaling properties of the 
invasion front of cancerous tumors may be described by those of the KPZ equation with a fractal dimension of $D_f
\approx 1.25\pm 0.05$ [32-34]. Analysis of two-dimensional (2D) interfacial spreading dynamics of quasilinear Vero 
cell colony fronts in methylcellulose-containing culture medium [35], as well as discrete models of growing cell 
populations [36] and tumor growth in uniform [37] and nonuniform culture medium [38] provided evidence that the 
scaling properties of the perimeter of tumors may be described by either the KPZ equation, or by a special limit of 
it in which the nonlinear term of the equation vanishes, and one has $\alpha\approx 0.63$ and $\beta\approx 0.75$. 
In addition, front instabilities have also been studied by various methods by considering physical interactions, as 
well as biomechanical effects [39-42].

Despite such detailed studies, one central question has remained unanswered: are the differences between the
universality classes indicated by distinct roughness and growth exponents due to the differences in the relevant 
biological processes and the physicasl factors that affect them, or are they simply a result of different methodologies 
and models? In other words, if we determine the properties of tumors' surface, will we be able to decipher the 
biological mechanism(s) that drive their growth?

One way of addressing the question is through studying the phenomenon by reaction-diffusion equations (RDEs) that, 
due to their ability for generating distinct patterns of reaction fronts, have been widely used [43] for modeling 
biological phenomena. Moreover, assuming that the diffusion coefficient that appears in the RDEs varies spatially has 
led to pattern formation in a multispecies model [44]. In addition, a recent study on a single-species RDE indicated 
that spatial fluctuations in the properties of cellular environments may lead to instabilities [45]. Use of the RDEs 
for modeling of tumor growth has a long history [46]. More recently, the RDEs have been used to study pattern formation 
and related issues at various scales, from molecular to tissue length scale [47-52].

In this paper we propose a model to study how the spatial variation of the local stiffness of the host tissue regulates 
tumor invasion. We consider stiffness as a regulator of cellular diffusion and assume that, due to the spatial 
fluctuation of the stiffness, the local diffusion coefficient also varies spatially. By solving a particular RDE in the 
presence of quenched disoder represented by a spatially-varying local diffusion coefficient, we attempt to imitate 
tumor growth and spreading in a heterogeneous environment. We then analyze the morphology of tumor invasion front as a 
function of the intensity of the fluctuations and the spatial correlations in the quenched disorder, as well as other 
relevant parameters. Our results indicate that the morphology of the invasion front is regulated by all such parameters,
as a result of which new universality classes emerge that are distinct from those of the KPZ and related models.

The rest of this paper is organized as follows. In the next section the proposed model is described, and the
numerical simulation of the governing equation is explained. The results are presented and discussed in Sec. III, 
where we study the effect of a variety of important factors on the shjape of tumor invasion front. A summary of the
main results of the paper is given in Sec. IV.

\begin{center}
{\bf II. THE MODEL}
\end{center}

Tumor growth commences with unlimited duplication, which later on is supported by the invasion into the surrounding 
tissues by its cells through diffusive or chemotaxic migration [53-55]. Though the dependence of cells' diffusion 
coefficient on the tissue stiffness is very complex and currently under investigation [9,56], for sufficiently stiff 
tissues, when tumor cells rely on proteolytic migration, their diffusivity is inversely regulated by stiffness 
[14,17]. If cellular migration faces stochastic local obstacles or is hindered in some direction, a change in the
diffusion coefficient for the entire environment [57] or in a specific direction [55] will be observed. Thus, the 
smaller motility may be viewed as a smaller diffusion coefficient. The exact nature of the dependence of the 
diffusivity on the stiffness has not been quantified yet, but it is clear that if the stiffness fluctuates locally, 
so would also the cells' diffusivity, because if the former is caused by some sort of spatial heterogeneity, so should
be the latter. 

Among the several theoretical approaches that have been proposed for modeling tumor growth, we consider a specific form
of the RDE, namely, the Kolmogorov-Petrovsky-Piskunov (KPP) equation [58], also known as the Fisher-KPP (FKPP) equation
[59], which is used frequently in studies of population dynamics. According to the FKPP equation, the cell density is 
governed by
\begin{equation}
\frac{\partial C}{\partial t}=R(1-C)C+\mbox{\boldmath$\nabla$}\cdot(D\mbox{\boldmath$\nabla$}C)\;,
\end{equation}
which describes the evolution of a population in a heterogeneous environment in which $R$ is the cells' duplication 
(reaction) rate, and $D$ is their local diffusion coefficient that varies spatially. If we discretize Eq. (4) by, 
for example, the finite-difference method, then, the cell density at each grid (or lattice) point $i$ is governed 
by [60],  
\begin{equation}
\frac{\partial C_i(t)}{\partial t}=\sum_{j\in\{i\}}D_{ij}[C_j(t)-C_i(t)]+RC_i(t)[1-C_i(t)]\;,
\end{equation}
where $\{i\}$ represents the set of nearest neighbours of grid point $i$, and $D_{ij}$ is the local diffusivity 
between $i$ and $j$. While the REDs with linear reactions in heterogeneous environments have been solved by 
approximate analytical approaches [61,62], the presence of the nonlinear reaction term in Eqs. (4) and (5) excludes
the possibility of applying the previous approaches to the present problem.  

The duplication rate $R$ of the cancer cells has been reported to be [63-65], $R\sim 10^{-5}$s$^{-1}$. The motility
of cancer cells {\it in vitro} has been reported to take on various values, ranging from [66] $D\sim 10^{-9}$ 
cm$^2$/s to [65] $D\sim 10^{-7}$ cm$^2$/s. Due to the microenvironmental effects [4] the diffusion coefficient
{\it in vivo} may be smaller. As such, we analyzed the range, $10^5\leq R/D\leq 10^8$, in order to include most
of the possible {\it in vivo} values.

To solve Eq. (5) numerically, we set the distance between two neighboring grid points at $100\;\mu$m. The rational 
behind such a grid block size is the fact that, in order to use a mean-field equation such as Eq. (5) for the cells' 
density, one needs to consider a mesoscopic length scale where the units - the grid blocks - contain a large number of 
cells. At the same time, the size of the grid blocks must be small compared to the overall size of the system under 
study. We used the forward-time central-space method that typically converges to the solution and is stable for the 
FKPP equation. To do so, for each grid site we solve for $C$ by writing 
\begin{displaymath}
C_i(t+\Delta t)=C_i(t)+\Delta t C_i(t)[1-C_i(t)]+\frac{\Delta t}{4\delta^2}\sum_{j\in\{i\}}D_{ij}[C_j(t)-C_i(t)]\;,
\end{displaymath}
where $\Delta t$ is the time step, and $\delta$ is the size of the grid blocks. Stability requires that 
$D\Delta t/(4\delta^2)<0.5$. The time step $\Delta t$ was to be 10 s or larger, depending on the $R/D$ ratio. We then 
solved the discretized equations on a computational grid of width $L$ with periodic boundary condition in the horizontal
direction in Fig. 1. The initial condition was $C_i=C_0$ for $(i,j)\leq n$; we typically used $n=6$, as shown in Fig. 
1(a). Since the number of cells is an integer, their density is quantized and small values of $C$ are not meaningful. 
Thus, one needs to define a cutoff $\epsilon$ for the density to represent the individuality of the species. Such a 
cutoff also affects the invasion velocity [67]. To determine an appropriate value for $\epsilon$, we carried out an 
analysis of the sensitvity of the results to $\epsilon$, which will be described shortly.

\begin{figure}
\centering
\includegraphics[width=0.26\linewidth]{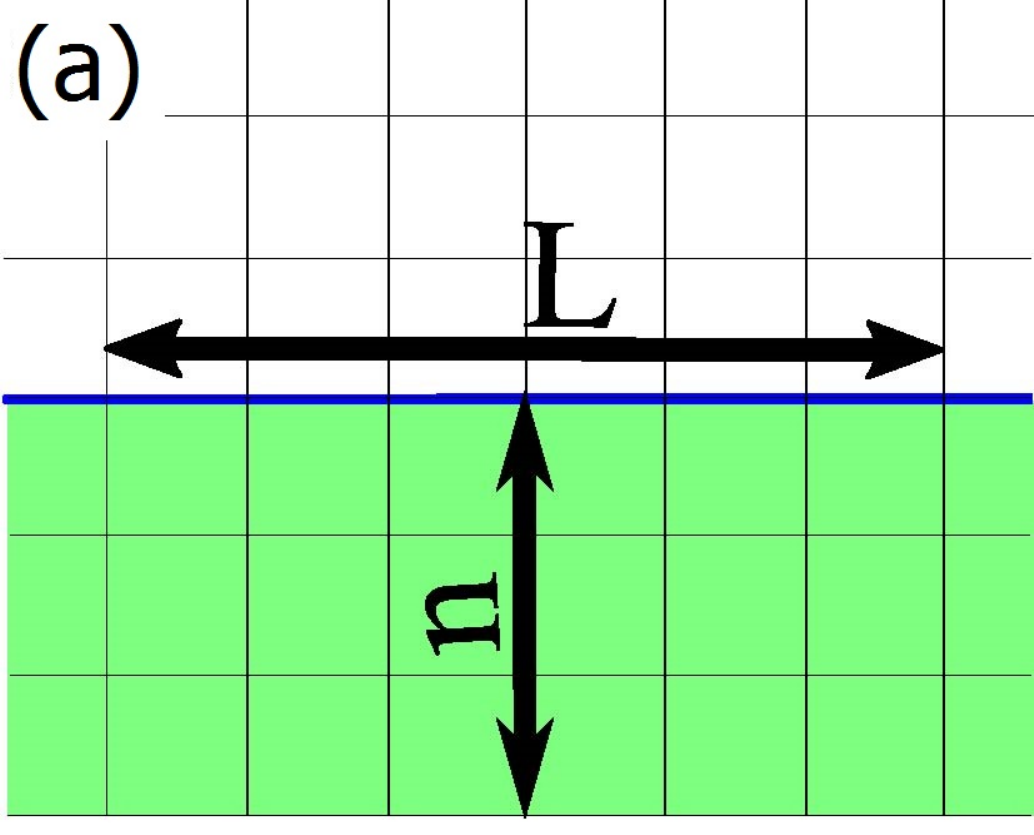}
\includegraphics[width=0.26\linewidth]{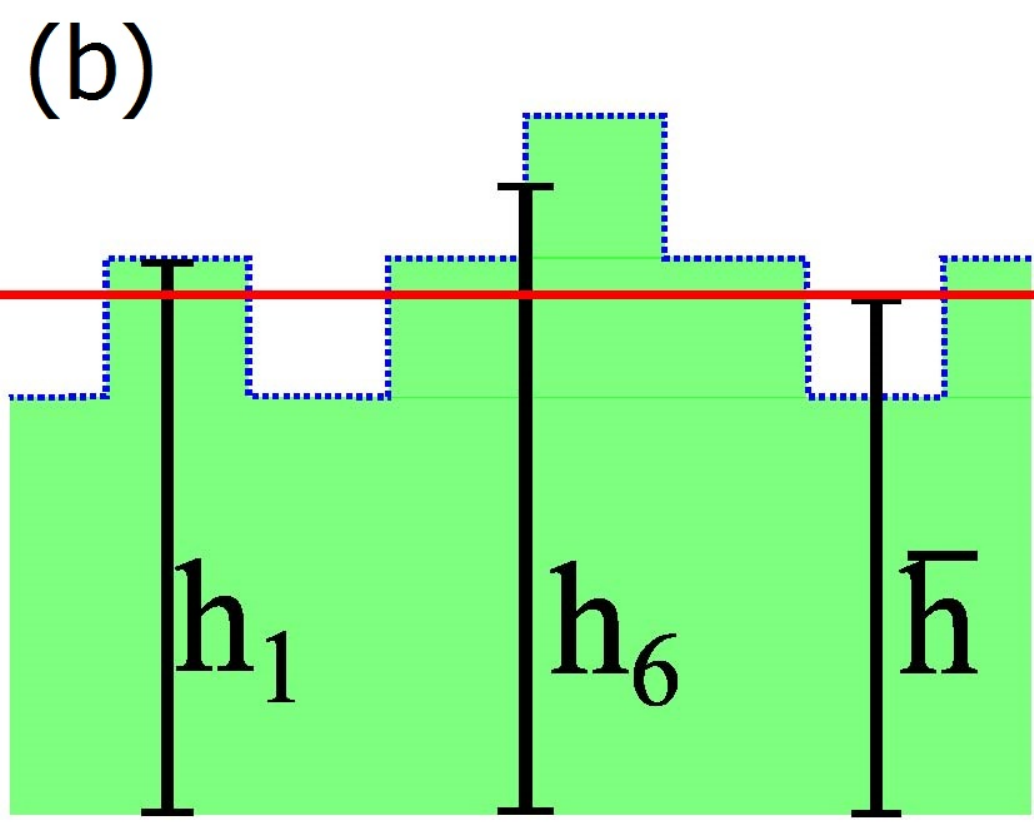} 
\includegraphics[width=0.48\linewidth]{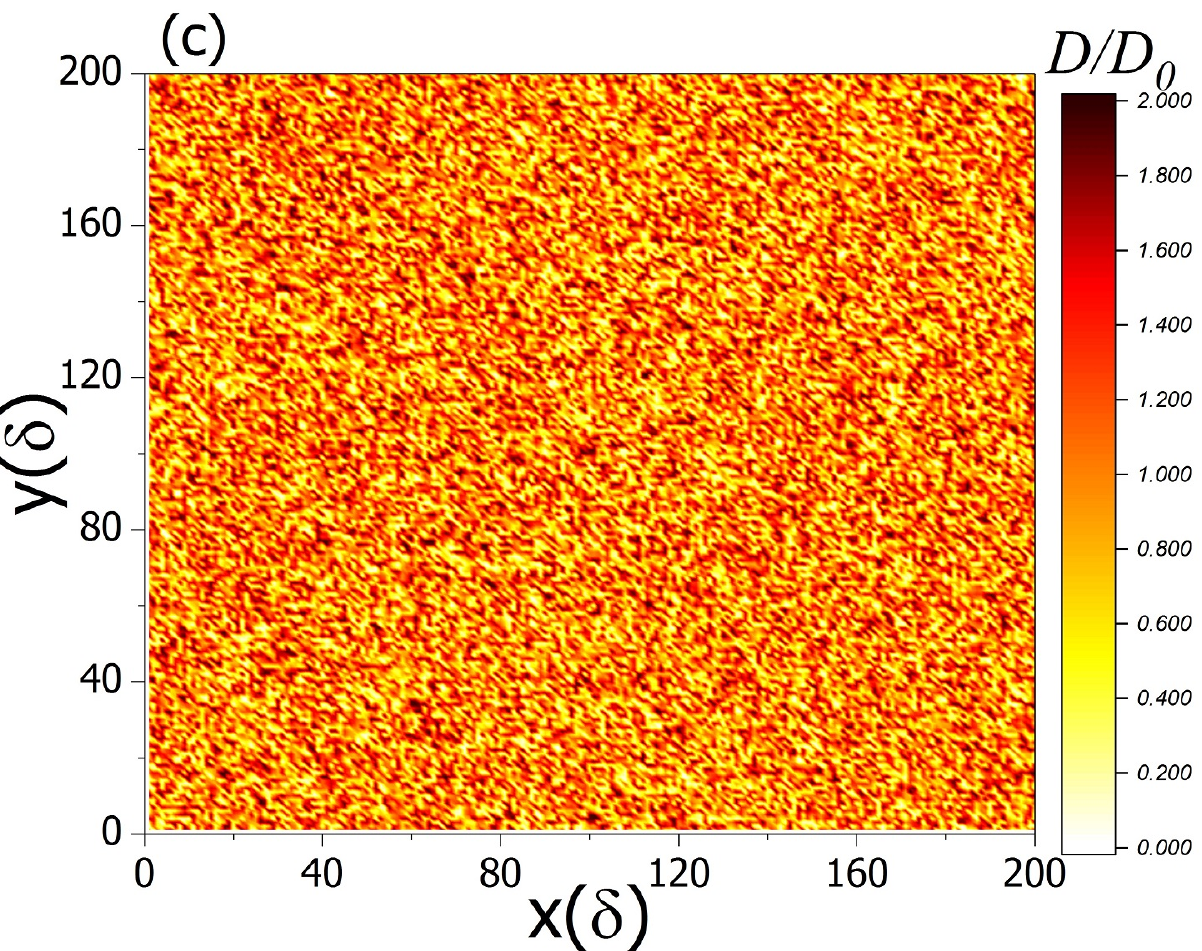}
\includegraphics[width=0.48\linewidth]{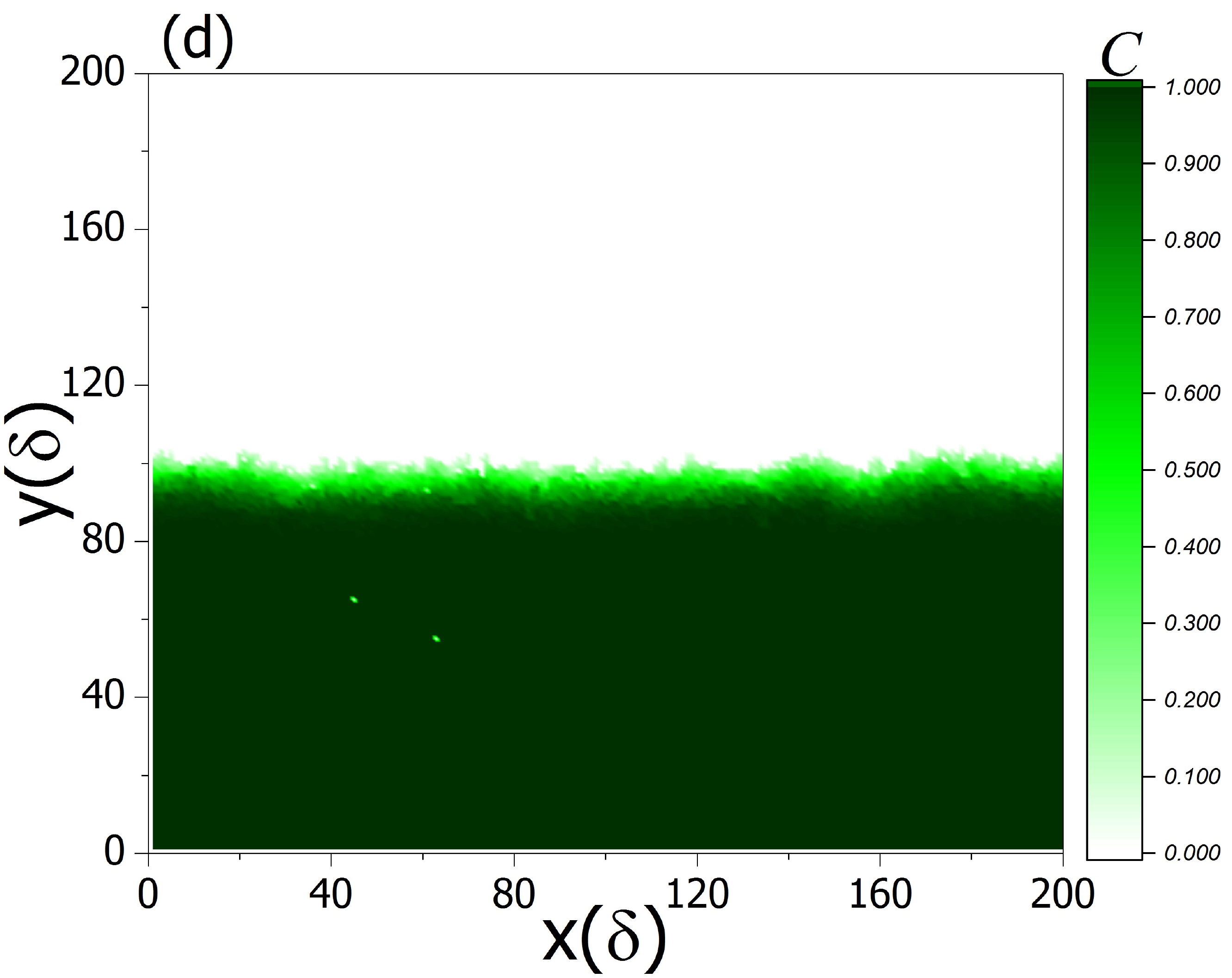}
\caption{Schematic illustration of the model. (a) The discretized medium with a grid block size of $\delta$. The
initial condition in the medium of size of $L$ is $C=C_0$ up to the width $n=3$. Periodic boundary condition was 
imposed in the horizontal direction. (b) The tumor's surface is defined as those edges that are around the border 
units, with each edge above the occupied units being the border edge, and with the border edges generating the straight
solid line. For a homogeneous cellular environment, the morphology of the interface remains unchanged and the solid 
line moves with a fixed velocity. (c) A realization of a heterogeneous cellular environment wit no spatial correlations
in the values of the diffusivity $D$ and the corresponding cell density with $R/\langle D\rangle=10^7$. (d) The 
distribution of the density $C$ corresponding to (c) after some time. Where density approaches zero is considered as 
the invasion front. Contrary to homogeneous cellular media, the invasion front deviates from linear geometry, giving 
rise to and average height $\bar{h}$. The initially straight line becomes ragged as time passes. Similar to the 
conventional approach in interface studies, we analyze the interface and quantify its dependence on the environmental 
fluctuations, initial conditions, and $R/D$.}
\end{figure} 

The simulations were carried out with up to 50,000 time steps. As $C_i(t)$ evolves over time, the surface, defined 
by the edges of border sites that are located between regions with zero and nonzero concentrations -dashed line in 
Fig. 1(b)- becomes rough. Each point on the surface between the two regions has a height of $h_i(t)$ with an average
height, $\bar{h}(t)=\sum_{i=1}^N h_i(t)/N$, where $N$ is the number of border edges. In a homogeneous medium the 
interface moves with a velocity $d\bar{h}/dt=2\sqrt{DR}$. To characterise the structure of the interface, we 
analyzed the surface width $W$, defined by Eq. (1), which provides a quantitative measure of the fluctuations at 
the interface. The average values of $W$ were computed for $10^3$ realizations. 

For a constant diffusion coefficient, i.e., uniform tissue stiffness, one has $W=0$ and, therefore, a constant $D$ 
that would not lead to instability of the invasion front. As mentioned earlier, however, tissues do not have a 
uniform stiffness. Analysis of host tissues by atomic force microscopes [18-21] has revealed huge fluctuations in 
the tissues' Young's modulus. To account for the effect of such fluctuations, we allowed the local diffusion 
coefficient, i.e., $D_{ij}$, to vary spatially. Thus, we considered heterogeneity over the same scale as the size of
the grid blocks and set, 
\begin{equation}
D_{ij}=D_0+\xi R_{ij}\;,
\end{equation}
where $R_{ij}$ varies according to a uniform distribution in $[-0.5,0.5]$, and $\xi$ represents the strength of the 
fluctuations. Figure 1(c) presents a realization of the heterogeneous environment with no spatial correlation. Such 
a heterogeneity leads to fluctuations in the invasion front , as Fig. 1(d) indicates. A main goal of the present 
paper is to study the dynamics of such irregular invasion fronts. To simulate a more realistic model, we assumed 
that spatial correlations exist between the stochastic values of the diffusivities; see below. 

To identify the appropriate cutoff $\epsilon$, we began the computations with a single-cell density threshold. Due 
to the way we have set up the model, the single-cell density is 0.01, which represents the upper limit for 
$\epsilon$. We then solved Eq. (5) with $\epsilon=10^{-2}$, and $C_0=0.2$, $n=6$, and $L=128$, where $L$ is defined 
in units of the grid blocks. We then decreased $\epsilon$ to study the dependence of the important properties on
$\epsilon$. As Fig. 2(a) indicates, the velocity of the front converges slowly with decreasing $\epsilon$. However, 
since our main concern is the geometry of the invasion front, we focused on the surface roughness and its width $W$ 
in order to analyze the convergence of the results with decreasing $\epsilon$. As Fig. 2(b) demonstrates, $W$ 
converges faster with decreasing $\epsilon$, such that even $\epsilon=10^{-3}$ is a good approximation, which is 
what we used in the rest of the computations.

\begin{figure}  
\centering
\includegraphics[width=0.48\linewidth]{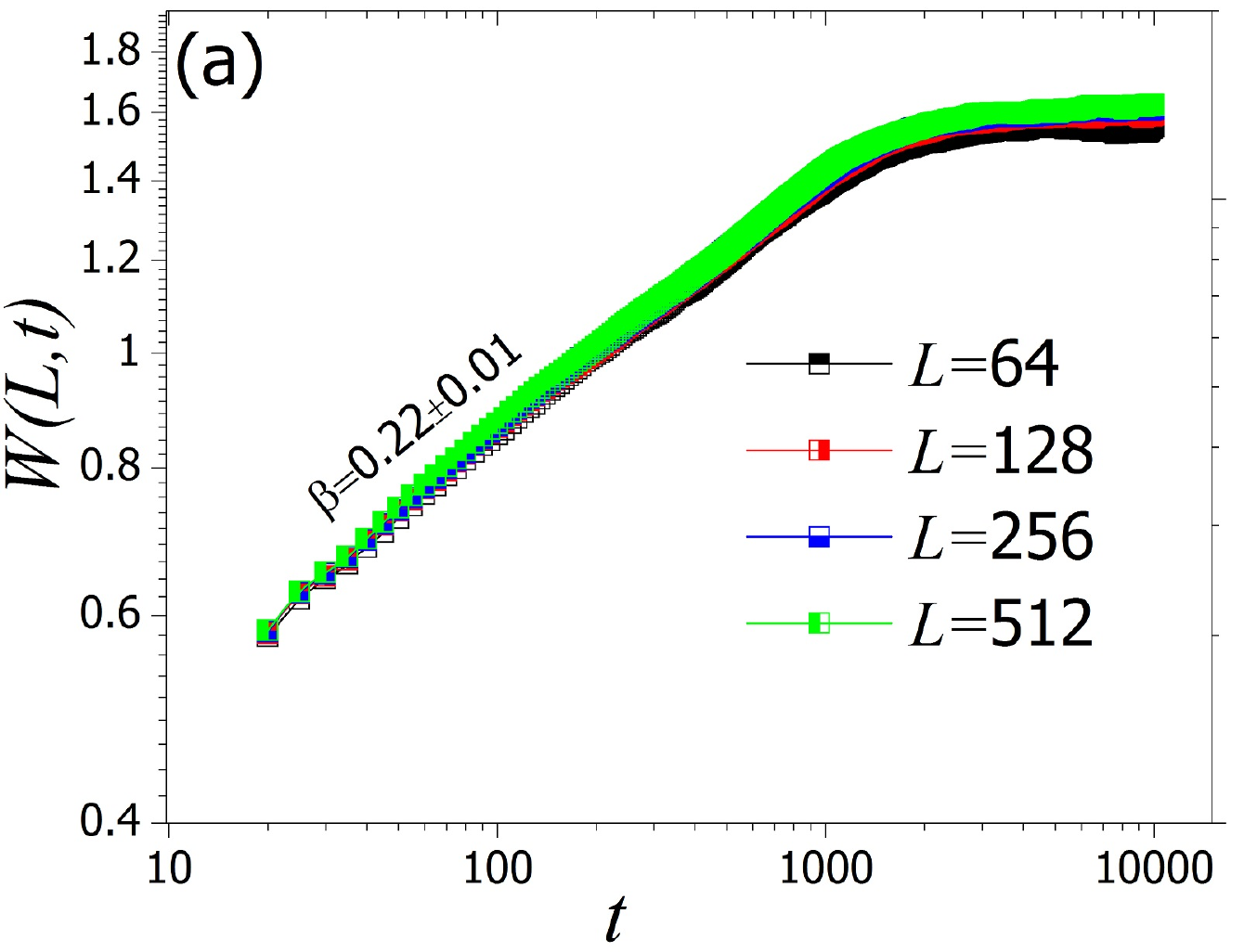}
\includegraphics[width=0.48\linewidth]{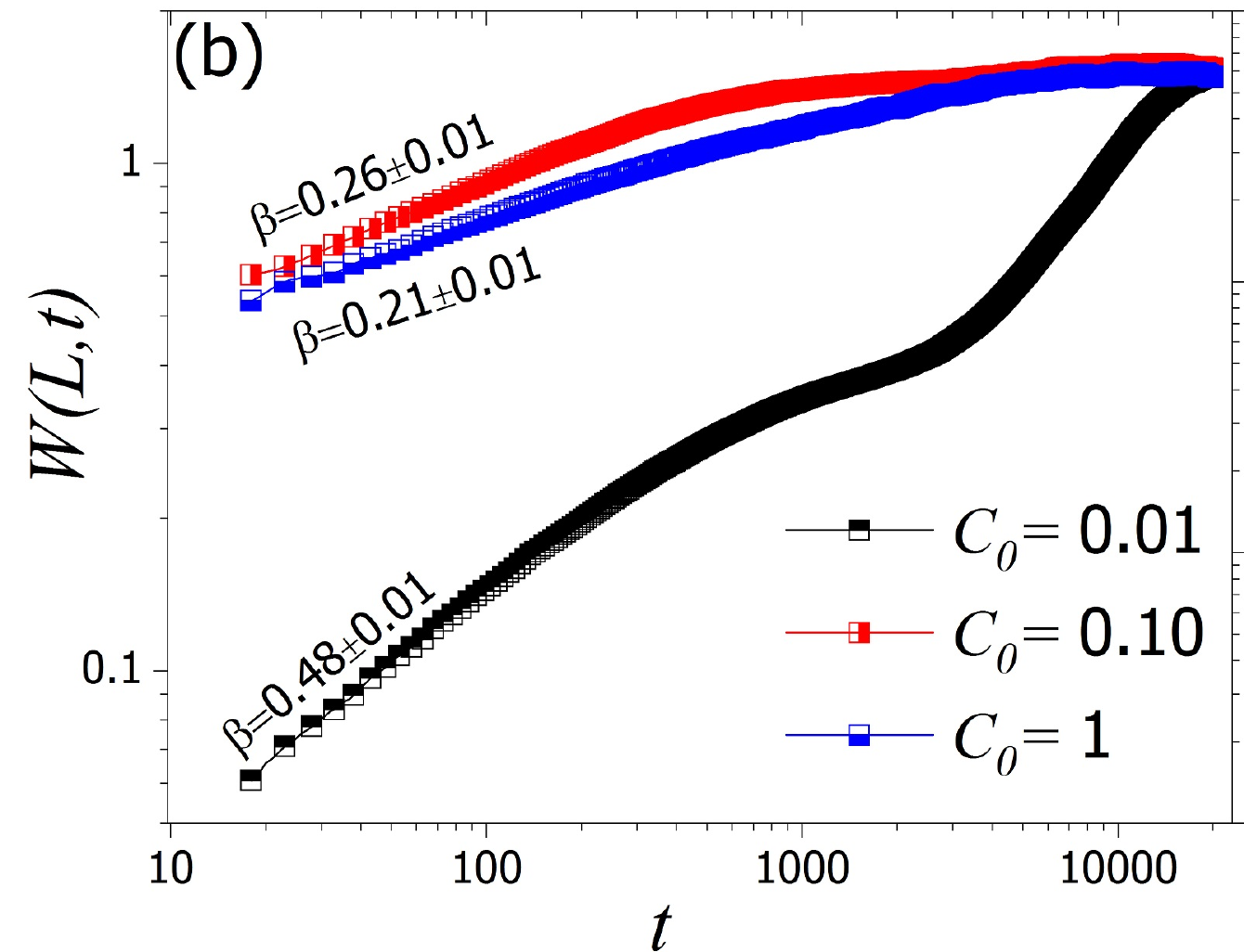}  
\includegraphics[width=0.48\linewidth]{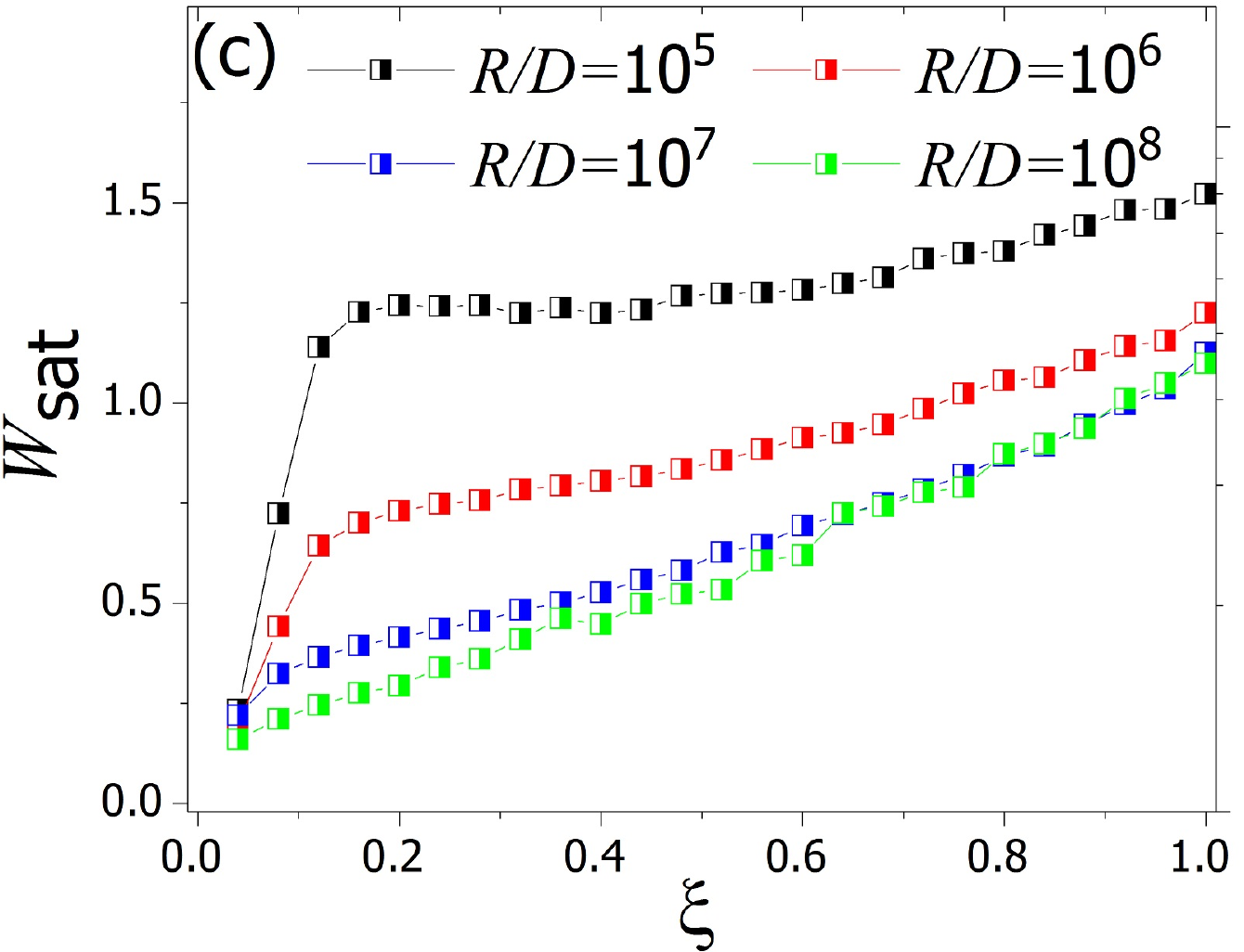}  
\includegraphics[width=0.48\linewidth]{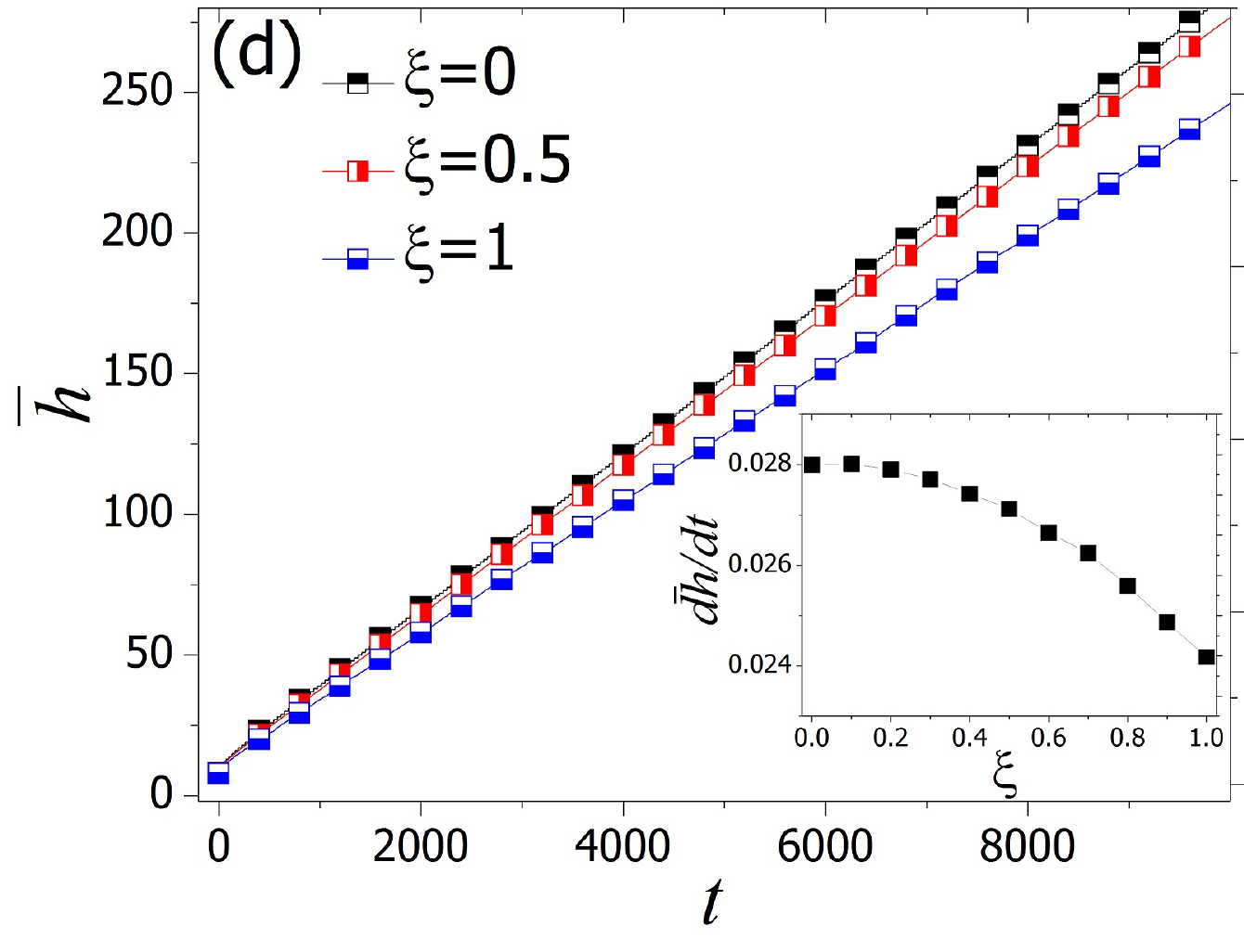}    
\caption{(a) Dynamic evolution of the width $W(L,t)$ for $\xi=1$, $C_0=0.2$, and $R/D =10^5$. The roughness exponent 
is, $\alpha=0$. (b) Scaling of $W(L,t)$ with time $t$ for $C_0=10^{-2},\;10^{-1}$ and 1. (c) $W_{\rm sat}$ versus the
intensity $\xi$ of the local diffusivity fluctuations. $W_{\rm sat}$ increases with increasing $\xi$ and decreases 
with increasing $R/D$. (d) Dynamic evolution of the mean interface height $\bar{h}$ versus $\xi$. The inset shows the
velocity of the invasion front.}
\end{figure}

\begin{center}
{\bf III. RESULTS}
\end{center}

As the first case study, we computed the dynamic evolution of the width $W(L,t)$ of the tumor invasion front for 
various sizes $L$ by solving Eq. (4) using Eq. (5) with the initial condition $C_0=0.2$ and $n=6$, where $n$ is 
defined by Fig. 1(a). As Fig. 3(a) indicates, $W(L,t)$ increases according to the power law (3) with $\beta\approx 
0.25\pm 0.01$. Once the width saturates, variation of $L$ does not affect the width, implying that the roughness 
exponent $\alpha=0$. 

If, however, we change the initial concentration $C_0$, the growth exponent $\beta$ turns out to vary with $C_0$. 
For example, Fig. 3(b) indicates that the initial concentrations, $C_0=0.01,\;0.1$ and 1 lead to $\beta=0.48
\pm 0.01,\;0.26\pm 0.01$ and $0.21\pm 0.01$, respectively. These results are not only interesting, but also in our 
view very important because they point to the initial condition as a key factor in the growth of tumors, which, 
to our knowledge, has been overlooked in previous experimental [31-34] and theoretical [35-37,68-73] studies. 
The apparent sensitivity to the initial conditions is presumably due to the nonlinear nature of the governing RDE 
that we solve. Thus, if the initial conditions in such studies had been varied, the exponent $\beta$ and, therefore,
 the universality class of the surface growth model could have been quite different from what they reported. Our 
analysis also indicates that $W_{\rm sat}$, the saturated width of the interface at steady state, remains unchanged.
Therefore, $W_{\rm sat}$ may be viewed as a reliable measure for assessing the properties of the interface. Using 
$W_{\rm sat}$ as an indicator, we study the effect of the various parameters, including the intensity of the 
fluctuations of the local diffusion coefficients, the normalized reaction rate $R/D$, and the type of the 
correlations in the spatial fluctuations, where $D$ is the average diffusion coefficient.

To understand how the spatial fluctuations lead to interface roughnening, we began with $\xi=0$, Eq. (6), and 
increased it to one, and studied how the roughness is regulated by the intensity of the fluctuations. As Fig. 3(c) 
indicates, for a fixed $R/D$ $W_{\rm sat}$ increases with $\xi$, whereas it decreases with increasing $R/D$ if 
$\xi$ is held constant. As such, we anticipate small (large) values of $R$ ($D$) lead to interface roughening. To 
quantify how $R/D$ regulates interface width at steady state, the dependence of $W_{\rm sat}$ of the ratio was 
computed. Figure 3(c) shows that increasing $R/D$ delays interface roughening. The reason that $R/D$ influences
the structure of the front is the fact that the reaction temporally relaxes the fluctuations, due to the 
spatial fluctuations in the diffusivity.

Another important property of invasion is its propagation velocity. While one can show analytically that the 
velocity attains a value of $2\sqrt{DR}$, retrieving this limiting value by numerical simulations must consider much
smaller values of $\epsilon$. As mentioned earlier, we fixed $\epsilon$ at $10^{-3}$ and, therefore, we analyzed the
dependence of the velocity with this $\epsilon$. As Fig. 3(d) shows, the average height grows slower when we 
increase $\xi$ and, while the invasion velocity decreases.

\begin{figure}
\centering
\includegraphics[width=0.48\linewidth]{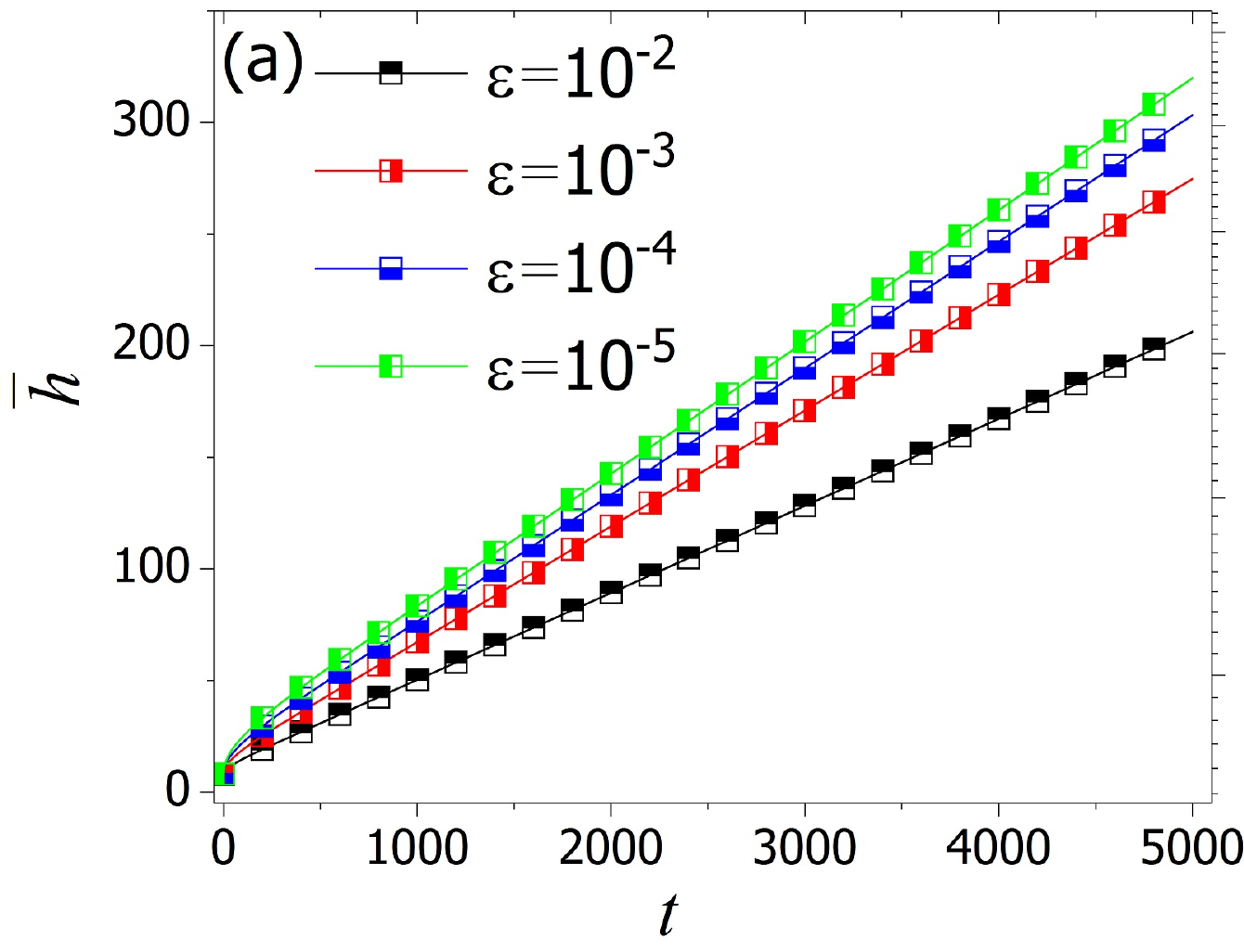}
\includegraphics[width=0.48\linewidth]{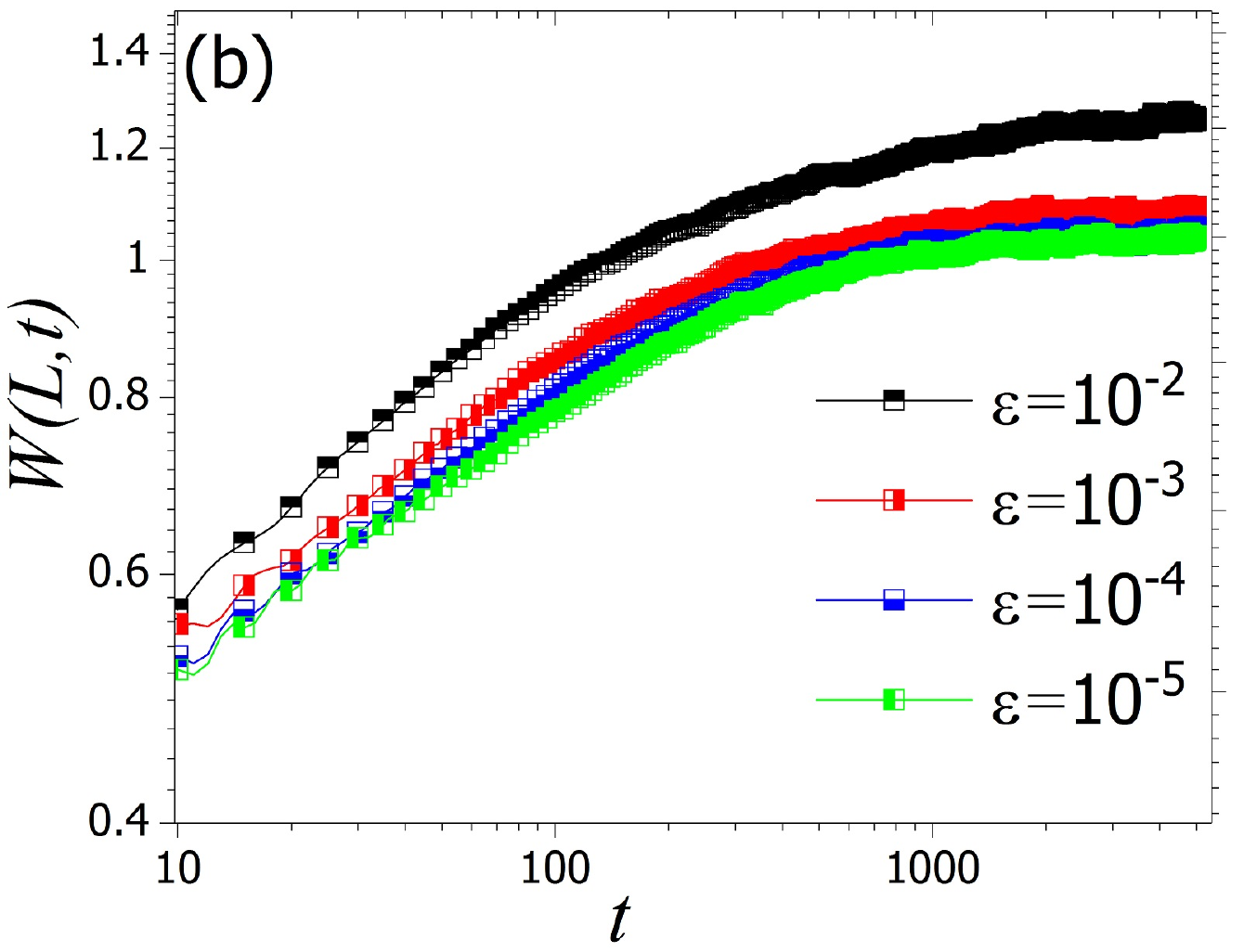}  
\caption{Analysis of the dependence of (a) the mean height $\bar{h}$ (and, hence, the velocity) and (b) the width 
$W(L,t)$ of the interface on the threshold $\epsilon$ with $C_0=1$, $n=6$ and $R/D=5\times 10^6$. While for invasion 
velocity one needs to consider a smaller values of $\epsilon$, the interface width converges faster.}
\end{figure}

We may then summarize the results presented in Fig. 3 as follows: (i) Biological surfaces do possess some of the 
scaling properties that the growing surfaces studied in other contexts have, but are not necessarily characterized 
by the well-known universality class of the KPZ and related equations, i.e., by the same numerical values of the
exponents $\alpha$ and $\beta$. (ii) The intensity of the stiffness fluctuations increases the width of the 
invasion front. (iii) Increasing $R/D$ decreases the surface width, and (iv) increasing the stochastic quantity 
$\xi$ decreases invasion velocity.

\begin{figure} 
\centering 
\includegraphics[width=0.65\linewidth]{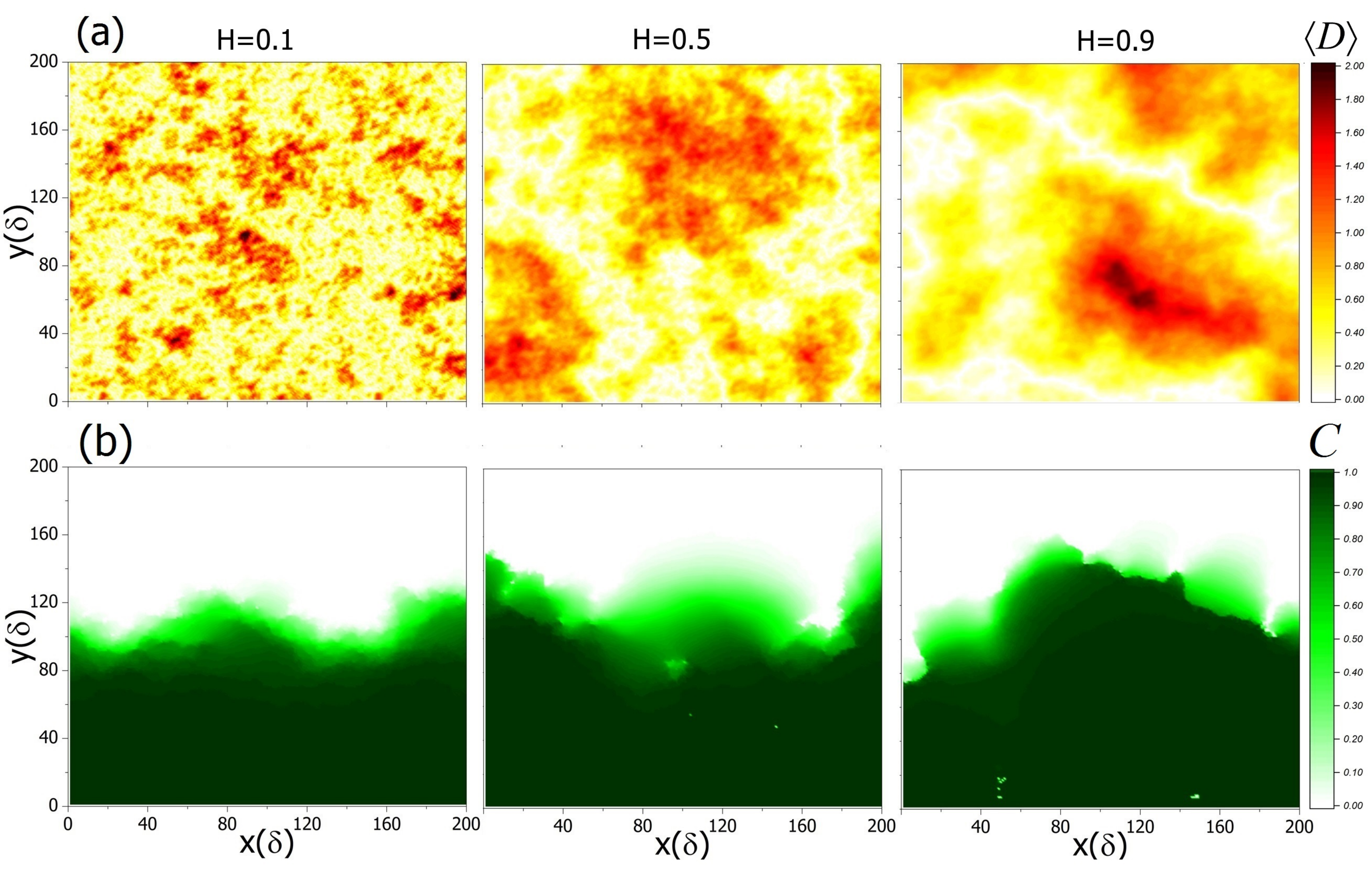}   
\includegraphics[width=0.38\linewidth]{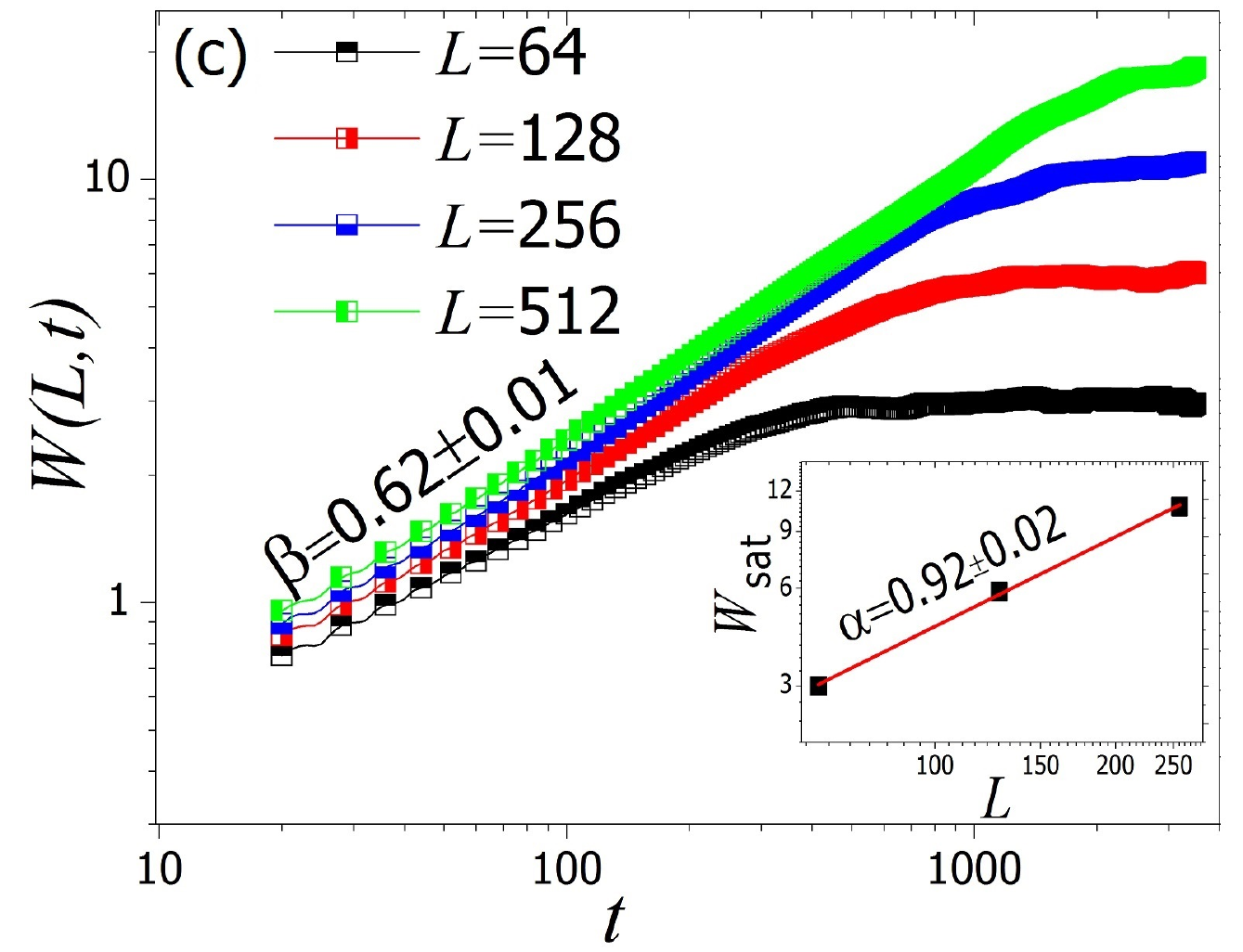}  
\includegraphics[width=0.38\linewidth]{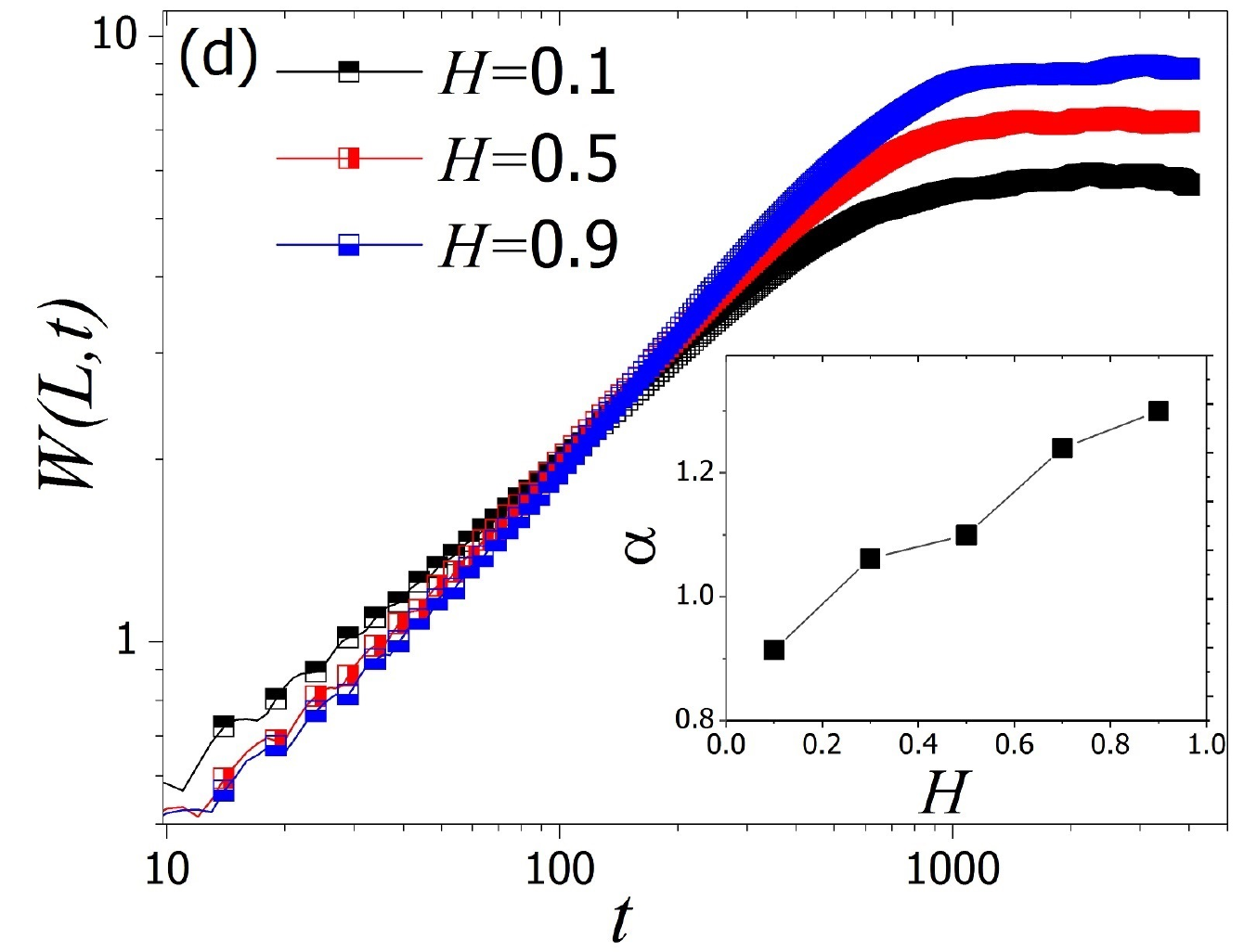} 
\includegraphics[width=0.38\linewidth]{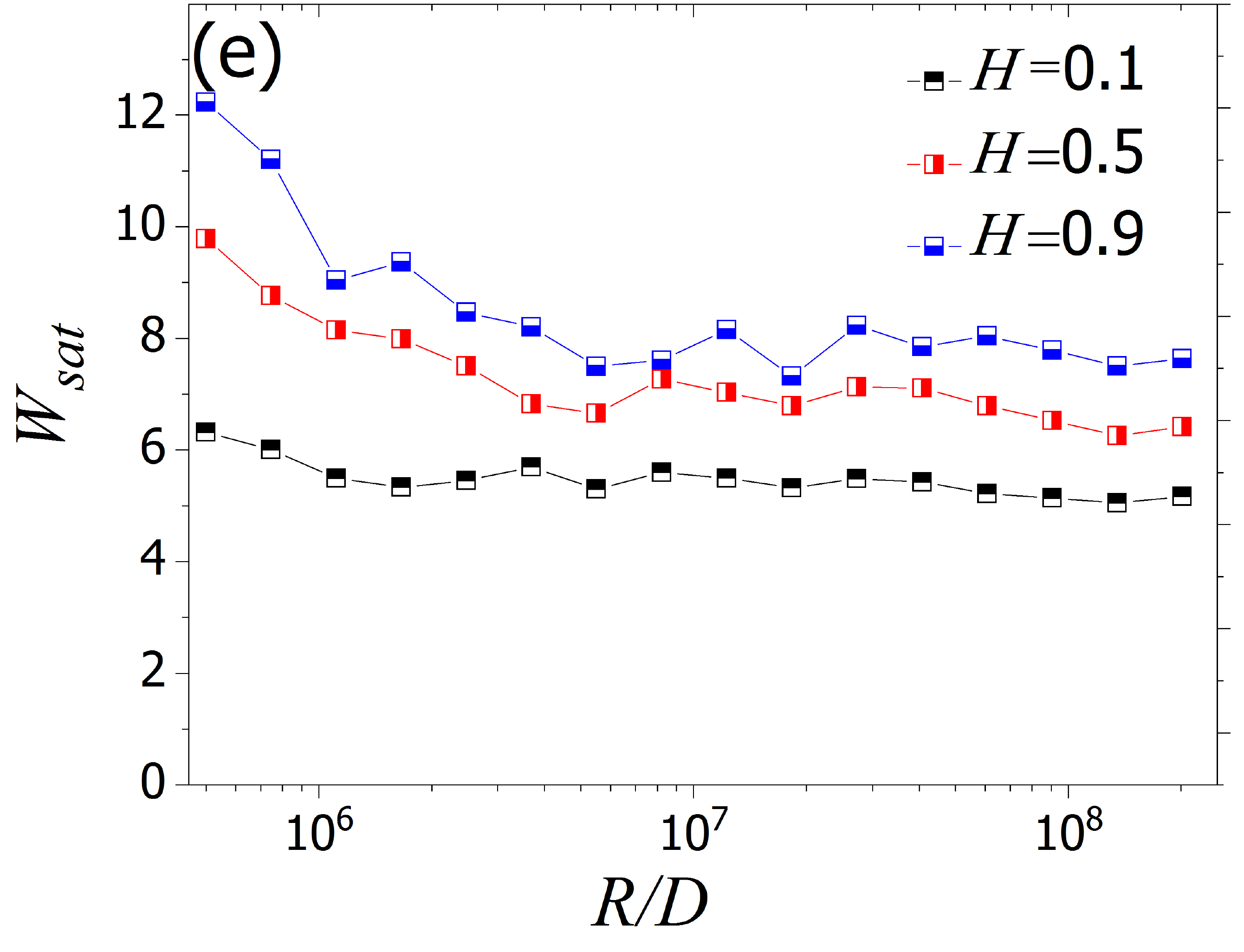} 
\caption{(a) Realizations for heterogeneous environments with $\langle D\rangle$ being the mean local diffusion coefficient, and $H$ the Hurst exponent. (b) The corresponding cell density for $R/D=2\times 10^6$, $D_0=0$ and the strength of the fluctuation of diffusivity $\xi=1$. (c) Scaling of the width $W(L,t)$ with time $t$ in a cellular medium with $H=0.1$, $R/D=10^8$, the initial concentration $C_0=1$, and $n=6$. The inset shows the dependence of the saturation width $W_{\rm sat}(L)$ on the linear size $L$ that leads to the roughness exponent $\alpha=0.92 \pm 0.01$. (d) Dynamic evolution of $W(L,t)$ for several values of $H$ and the same initial condition. The inset shows thedependence of $\alpha$ on $H$. (e) Dependence of $W_{\rm sat}(L)$ on $R/D$ and $H$ for $L=64$.}
\end{figure}

As mentioned earlier, the spatial distribution of the tissues' stiffness appears to be correlated [20,21], rather 
than being completely random. Therefore, the local diffusivities should also be spatially correlated. The apparent 
correlation function has not been quantified yet, however. At the same time, long-range/extended correlations are 
prevalent in many natural phenomena [74-76], including in biological systems [77,78], and are accurately represented
by the fractional Brownian motion [79] (FBM). If we consider two points at {\bf x} and {\bf x'}, the FBM is defined 
by,
\begin{equation}
\langle(R_{\bf x}-R_{\bf x'})^2\rangle\sim ({\bf x}-{\bf x'})^{2H}\;, 
\end{equation}
with $H$ being the Hurst exponent such that $H>0.5$ ($H<0.5$) induces positive (negative) correlations, and $H=0.5$ 
corresponds to completely random successive increments. Figure 4(a) presents the morphologies that the FBM generates
for various values of $H$.

Thus, we used the FBM to generate spatially-correlated local diffusivities. A $1024\times 1024$ computational grid 
with a FBM surface was generated. Then, we used Eq. (6) with $R_{ij}=(|R_i+R_j|)/2$, where $R_i$ and $R_j$ are the 
FBM-generated values at locations (grid sites) $i$ and $j$, with $j$ being a nearest neighbor of $i$. We set $\xi$ 
such that $\langle R_{ij}\rangle=0.5$. In this way, the FBM provided us with a stochastic function that generates 
spatially-correlated fluctuations in the local diffusivities that are regulated by the Hurst exponent $H$. Note that
we do not claim at this time that the FBM characterizes the true nature of the correlations in biological system
that we study, rather we use it as a typical stochastic process that generates extended correlations that have been 
identified in other biological processes and heterogeneous media.

The effect of the spatial correlations, which are part of the quenched disorder in the morphology of the cellular 
medium, is independent of the effect of the reaction. Thus, the scaling properties of the tumor front depend on the 
initial condition. The ratio $R/D$ has a ``softening'' effect, as the reaction slows down the diffusion. Hence, we 
concentrated on the effect of the spatial correlations on the structure of the invasion front by computing the 
dependence of $W_{\rm sat}$ on the Hurst exponent $H$, in order to analyze the properties of the perimeter of a 
growing tumor. Figure 4(b) presents examples of the rough fronts and their dependence on the Hurst exponent $H$.

Next, consider, the limit $H=0.5$, and let the surface evolve from the initial condition, $C_0=0.2$ with $n=6$ and 
$R/D=10^5$. As Fig. 4(c) indicatess, the scaling of the width $W(L,t)$ follows Eq. (3) with $\beta\approx 0.62\pm 
0.01$. Moreover, similar to the case of no correlations, $\beta$ depends on the initial condition. Irrespective of 
the initial conditions, however, one has, $W_{\rm sat}(L)\propto L^\alpha$ with the roughness exponent being,
$\alpha\approx 0.9$; that is, $\alpha$ is independent of the initial condition, whereas the growth exponent $\beta$ 
is does depend on $R/D$. 

The dependence of $W_{\rm sat}(L)$ on the Hurst exponent $H$ is shown in Fig. 4(d). It indicates that with 
increasing $H$, i.e., with the diffusivities becoming more positively correlated, the width $W_{\rm sat}$ of the 
tumor's perimeter at steady state also increases. The inset of Fig. 4(d) presents the dependence of $\alpha$ on $H$,
indicating relatively strong dependence of $\alpha$ on $H$.

\begin{center}
{\bf IV. SUMMARY}
\end{center}

Using a well-known approach for the study of population dynamics, we investigated tumor growth in heterogeneous 
cellular environments. By treating the local diffusion coefficient, a measure of the ability of the cells to 
advance, as a spatially-varying quantity, we studied the morphology of the tumor invasion surface in heterogeneous 
cellular environments. Our results indicate that the previous classifications of biological interfaces in terms of 
their scaling properties should be reconsidered carefully, because the effect of the initial conditions had been 
overlooked, whereas, as our study indicated, the scaling exponents that characterize the structure of the tumor 
surface manifest great sensitivity to the initial conditions. 

We also showed that $W_{\rm sat}$, the saturation width of the tumor's surface, is an important characteristic 
quantity to study, in order to understand the effect of the heterogeneity in the cellular environment. Moreover, 
the effect of $R/D$, the ratio of the cells' multiplication rate $R$ and the average diffusivity $D$, on 
$W_{\rm sat}$ was shown to be important, as was the effect of the extended correlations in the local diffusion 
coefficient. Our results cast doubt on the classification of biological interfaces through well-known universality 
classes of surface growth processes, due to the effect of the heterogeneity and extended correlations, heretofore 
ignored. 

The sensitivity of nonlinear processes, such as the reaction-diffusion phenomena that we study, to the initial 
conditions appears to be a general feature of such phenomena, which has also been reported in other contexts. For 
example the universality of fluctuations in the KPZ model is sensitive to the initial conditions, i.e., whether
the initial substrate is wedge-like or flat at time $t=0$ [80-84]. In addition, we showed in a recent paper [52]
that the initial condition for the available nutrients strongly affects the statistics of tumor front. All such
results are consistent with what we present in this paper and, therefore, provide further evidence the initial
conditions play a fundamental role in the nonlinear processes in biological environments.

The model can be extended and improved. For example, During the invasion process, the migrating cells can dynamically
change their microenvironment, by either degrading the extra-cellular matrix (ECM) fibers to generate space for 
migration, or by forming stressed fiber bundles to propagate active forces. On large time scales, the expansion of the 
primary tumor mass would eventually generate pressures, which in turn stiffens the surrounding ECM. In addition, the
diffusivity can depend on the local concentration. Thus, we carried out a series of preliminary simulations in which
we assumed that, $D=D_0+\gamma C$, where $\gamma$ could be positive or negative. This did not, however, affect the 
invasion geometry significantly. WE hope to address such issues in the near future.

\begin{center}
{\bf ACKNOWLEDGMENT}
\end{center}
A.A.S. would like to acknowledge the supports from the Alexander von Humboldt Foundation and the research council of the University of Tehran.
We would like to thank L. Preziosi for reading the first draft of the paper and his useful comments.

\bigskip

\noindent $^\dagger$ab.saberi@ut.ac.ir

\newcounter{bean}
\begin{list}%
{[\arabic{bean}]}{\usecounter{bean}\setlength{\rightmargin}{\leftmargin}}

\item D. Wirtz, K. Konstantopoulos, and P.C. Searson, The physics of cancer: the role of physical interactions 
and mechanical forces in metastasis, Nat. Rev. Cancer {\bf 11}, 512 (2011).

\item F. Michor, J. Liphardt, M. Ferrari, and J. Widom, What does physics have to do with cancer? Nat. Rev. Cancer 
{\bf 11}, 657 (2011).

\item G. Charras and E. Sahai, Physical influences of the extracellular environment on cell migration, Nat. Rev. 
Mol. Cell Biol. {\bf 15}, 813 (2014).

\item C.D. Paul, P. Mistriotis, and K. Konstantopoulos, Cancer cell motility: lessons from migration in confined
spaces, Nat. Rev. Cancer {\bf 17}, 131 (2017).

\item T.A. Ulrich, E.M. de Juan Pardo, and S. Kumar, The mechanical rigidity of the extracellular matrix 
regulates the structure, motility, and proliferation of glioma cells, Cancer Res. {\bf 69}, 4167 (2009).

\item M. Cavo, M. Fato, L. Pe\~nuela, F. Beltrame, R. Raiteri, and S. Scaglione, Microenvironment complexity 
and matrix stiffness regulate breast cancer cell activity in a 3D in vitro model, Sci. Rep. {\bf 6}, 35367 (2016).

\item D.E. Discher, P. Janmey, and Y.-L. Wang, Tissue cells feel and respond to the stiffness of their substrate, 
Science {\bf 310}, 1139 (2005).

\item S.E. Reid, E.J. Kay, L.J. Neilson, A.-T. Henze, J. Serneels, E.J. McGhee, S. Dhayade, C. Nixon, J.B. Mackey, 
A. Santi, et al., Tumor matrix stiffness promotes metastatic cancer cell interaction with the endothelium, EMBO J. 
{\bf 36}, 2373 (2017).

\item A. Pathak and S. Kumar, Independent regulation of tumor cell migration by matrix stiffness and confinement, 
Proc. Natl. Acad. Sci. USA {\bf 109}, 10334 (2012).

\item V. Gkretsi and T. Stylianopoulos, Cell adhesion and matrix stiffness: coordinating cancer cell invasion and 
metastasis, Front. Oncol. {\bf 8}, 145 (2018).

\item R.G. Wells, The role of matrix stiffness in regulating cell behavior, Hepatology {\bf 47}, 1394 (2008).

\item S.S. Rao, S. Bentil, J. DeJesus, J. Larison, A. Hissong, R. Dupaix, A. Sarkar, and J.O. Winter, Inherent 
interfacial mechanical gradients in 3D hydrogels influence tumor cell behaviors, PLoS One {\bf 7}, e35852 (2012).

\item S.N. Kim {\it et al.}, ECM stiffness regulates glial migration in Drosophila and mammalian glioma models, 
Development {\bf 141}, 3233 (2014).

\item M.H. Zaman {\it et al.}, Migration of tumor cells in 3D matrices is governed by matrix stiffness along with 
cell-matrix adhesion and proteolysis, Proc. Natl. Acad. Sci. USA {\bf 103}, 10889 (2006).

\item T.A. Ulrich, A. Jain, K. Tanner, J.L. MacKay, and S. Kumar, Probing cellular mechanobiology in 
three-dimensional culture with collagen-agarose matrices, Biomater. {\bf 31}, 1875 (2010).

\item J. Fenner, A.C. Stacer, F. Winterroth, T.D. Johnson, K.E. Luker, and G.D. Luker, Macroscopic stiffness of 
breast tumors predicts metastasis, Sci. Rep. {\bf 4}, 5512 (2014).

\item K.M.A. Yong, Z. Li, S.D. Merajver, and J. Fu, Tracking the tumor invasion front using long-term fluidic 
tumoroid culture, Sci. Rep. {\bf 7}, 10784 (2017).

\item Y. Jamin {\it et al.}, Exploring the biomechanical properties of brain Malignancies and their pathologic 
determinants {\it in vivo} with magnetic resonance elastography, Cancer Res. {\bf 75}, 1216 (2015).

\item L. Xu, Y. Lin, J. Han, Z. Xi, H. Shen, and P. Gao, Magnetic resonance elastography of brain tumors: 
preliminary results, Acta Radiol. {\bf 48}, 327 (2007).

\item A.J. McKenzie, S.R. Hicks, K.V. Svec, H. Naughton, Z.L. Edmunds, and A.K. Howe, The mechanical microenvironment 
regulates ovarian cancer cell morphology, migration, and spheroid disaggregation, Sci. Rep. {\bf 8}, 7228 (2018).

\item M. Plodinec {\it et al.}, The nanomechanical signature of breast cancer, Nat. Nanotechnol. {\bf 7}, 757 (2012).

\item N. Bouchonville, M. Meyer, C. Gaude, E. Gay, D. Ratel, and A. Nicolas, AFM mapping of the elastic properties 
of brain tissue reveals kPa $\mu$m$^{−1}$ gradients of rigidity, Soft Matter {\bf 12}, 6232 (2016).

\item K. Schregel, N. Nazari, M.O. Nowicki, M. Palotai, S.E. Lawler, R. Sinkus, P.E. Barbone, and S. Patz,
Characterization of glioblastoma in an orthotopic mouse model with magnetic resonance elastography, NMR Biomed. 
e3840 (2017).

\item D. Fovargue, D. Nordsletten, and R. Sinkus, Stiffness reconstruction methods for MR elastography, NMR Biomed.
e3935 (2018).

\item A. Br\'u, J.M. Pastor, I. Fernaud, I. Br\'u, S. Melle, and C. Berenguer, Super-rough dynamics on tumor growth, 
Phys. Rev. Lett. {\bf 81}, 4008 (1998).

\item A. Br\'u, S. Albertos, J.L. Subiza, J.L. García-Asenjo, and I. Br\'u, The universal dynamics of tumor growth,
Biophys. J. {\bf 85}, 2948 (2003).

\item E. Claridge, P. Hall, M. Keefe, and J. Allen, Shape analysis for classification of malignant melanoma, J. Biomed. 
Eng. {\bf 14}, 229 (1992).

\item T.K. Lee and E. Claridge, Predictive power of irregular border shapes for malignant melanomas, Skin Res. Technol. 
{\bf 11}, 1 (2005).

\item J. P\'rez-Beteta {\it et al.}, Tumor surface regularity at MR imaging predicts survival and response to surgery 
in patients with glioblastoma, Radiology {\bf 288}, 218 (2018).

\item J. P\'erez-Beteta {\it et al.}, Morphological MRI-based features provide pretreatment and post-surgery survival 
prediction in glioblastoma, Europ. Radiol. {\bf 29}, 1968 (2018).

\item M. Kardar, G. Parisi, and Y.C. Zhang, Dynamic scaling of growing interfaces, Phys. Rev. Lett. {\bf 56}, 889 
(1986).

\item M.A.C. Huergo, M.A. Pasquale, A.E. Bolz\'an, A.J. Arvia, and P.H. Gonz\'alez, Morphology and dynamic scaling 
analysis of cell colonies with linear growth fronts, Phys. Rev. E {\bf 82}, 031903 (2010).

\item M.A.C. Huergo, M.A. Pasquale, P.H. Gonz\'alez, A.E. Bolz\'an, and A.J. Arvia, Dynamics and morphology 
characteristics of cell colonies with radially spreading growth fronts, Phys. Rev. E {\bf 84}, 021917 (2011).

\item M.A.C. Huergo, M.A. Pasquale, P.H. Gonz\'alez, A.E. Bolz\'an, and A.J. Arvia, Growth dynamics of cancer 
cell colonies and their comparison with noncancerous cells, Phys. Rev. E {\bf 85}, 011918 (2012).

\item M.A.C. Huergo, N.E. Muzzio, M.A. Pasquale, P.H. Pedro Gonz\'alez, A.E. Bolz\'an, and A.J. Arvia, Dynamic 
scaling analysis of two-dimensional cell colony fronts in a gel medium: A biological system approaching a quenched 
Kardar-Parisi-Zhang universality, Phys. Rev. E {\bf 90}, 022706 (2014).

\item N. Block, E. Sch\"oll, and D. Drasdo, Classifying the expansion kinetics and critical surface dynamics of 
growing cell populations, Phys. Rev. Lett. {\bf 99}, 248101 (2007).

\item B. Moglia, N. Guisoni, and E.V. Albano, Interfacial properties in a discrete model for tumor growth, Phys. 
Rev. E {\bf 87}, 032713 (2013).

\item B. Moglia, E.V. Albano, and N. Guisoni, Pinning-depinning transition in a stochastic growth model for the 
evolution of cell colony fronts in a disordered medium, Phys. Rev. E {\bf 94}, 052139 (2016).

\item P. Ciarletta, Buckling instability in growing tumor spheroids, Phys. Rev. Lett. {\bf 110}, 158102 (2013).

\item J.S. Lowengrub, H.B. Frieboes, F. Jin, Y.-L. Chuang, X. Li, P. Macklin, S.M. Wise, and V. Cristini, Nonlinear 
modelling of cancer: bridging the gap between cells and tumours, Nonlinearity {\bf 23}, R1 (2009).

\item M. Castro, C. Molina-Pa\'ris, and T.S. Deisboeck, Tumor growth instability and the onset of invasion, Phys. 
Rev. E {\bf 72}, 041907 (2005).

\item M.B. Amar, C. Chatelain, and P. Ciarletta, Contour instabilities in early tumor growth models, Phys. Rev. 
Lett. {\bf 106}, 148101 (2011).

\item S. Kondo and T. Miura, Reaction-diffusion model as a framework for understanding biological pattern 
formation, Science {\bf 329}, 1616 (2010).

\item P.K. Maini, D.L. Benson, and J.A. Sherratt, Pattern formation in reaction-diffusion models with spatially 
inhomogeneous diffusion coefficients, Math. Med. Biol. {\bf 9}, 197 (1992).

\item S. Paul, S. Ghosh, and D.S. Ray, Reaction–diffusion systems with fluctuating diffusivity; spatio-temporal chaos 
and phase separation, J. Stat. Mech., 033205 (2018).

\item R.A. Gatenby and E.T. Gawlinski, A Reaction-diffusion model of cancer invasion, Cancer Res. {\bf 56}, 5745 (1996).

\item Z. Dai and J.W. Locasale, Metabolic pattern formation in the tumor microenvironment, Mol. Syst. Biol.
{\bf 13}, 915 (2017).

\item Q. Zheng and J. Shen, Dynamics and pattern formation in a cancer network with diffusion, Commun. Nonlinear 
Sci. Numer. Simul. {\bf 27}, 93 (2015).

\item H. Youssefpour, X. Li, A. Lander, and J. Lowengrub, Multispecies model of cell lineages and feedback control in 
solid tumors, J. Theor. Biol. {\bf 304}, 39 (2012).

\item P. Kundr\'at and W. Friedland, Enhanced release of primary signals may render intercellular signalling ineffective
 due to spatial aspects, Sci. Rep. {\bf 6}, 33214 (2016).

\item M. Lee, G.T. Chen, E. Puttock, K. Wang, R.A. Edwards, M.L. Waterman, and J. Lowengrub, Mathematical modeling 
links Wnt signaling to emergent patterns of metabolism in colon cancer, Mol. Syst. Biol. {\bf 13}, 912 (2017).

\item Y. Azimzade, A.A. Saberi, and M. Sahimi, Role of the interplay between the internal and external conditions
in invasive behavior of tumors, Sci. Rep. {\bf 8}, 5968 (2018).

\item D. Hanahan and R.A. Weinberg, Hallmarks of cancer: the next generation, Cell {\bf 144}, 646 (2011).

\item E.T. Roussos, J.S. Condeelis, and A. Patsialou, Chemotaxis in cancer, Nat. Rev. Cancer {\bf 11}, 573 (2011).

\item Y. Azimzade, A.A. Saberi, and M. Sahimi, Regulation of migration of chemotactic tumor cells by the spatial
distribution of collagen fiber orientation, Phys. Rev. E {\bf 99}, 062414 (2019).

\item F. Kai, H. Laklai, and V.M. Weaver, Force matters: biomechanical regulation of cell invasion and migration 
in disease, Trends Cell Biol. {\bf 26}, 486 (2016).

\item Y. Azimzade aqnd A. Mashaghi, Search efficiency of biased migration towards stationary or moving targets in
heterogeneously structured environments, Phys. Rev. E {\bf 96}, 062415 (2017).

\item  A. Kolmogorov, I. Petrovskii, and N. Piskunov, A study of the diffusion equation with increase in the amount of 
substance and its application to a biological problem, in {\it Selected Works of A.N. Kolmogorov I}, edited by V.M. 
Tikhomirov (Kluwer, Amsterdam, 1991), p. 248; the translation of the original paper was by V. M. Volosov, Bull. Moscow 
Univ. Math. Mech. {\bf 1}, 1 (1937).

\item R.A. Fisher, The wave of advance of advantageous genes, Annal. Eugenics. {\bf 7}, 353 (1937).

\item M. Sahimi, B.D. Hughes, L.E. Scriven, and H.T. Davis, Stochastic transport in disordered systems, J. Chem. Phys. 
{\bf 78}, 6849 (1983).

\item M. Sahimi, Diffusion-controlled reactions in disordered porous media. I: Uniform distribution of reactants,
Chem. Eng. Sci. {\bf 43}, 2981 (1988).

\item R. Mojaradi and M. Sahimi, Diffusion-controlled reactions in porous media. II: Non-uniform distribution of 
reactants, Chem. Eng. Sci. {\bf 43}, 2995 (1988).

\item N.K. Haass, K.A. Beaumont, D.S. Hill, A. Anfosso, P. Mrass, M.A. Munoz, I. Kinjyo, and W. Weninger, Real-time 
cell cycle imaging during melanoma growth, invasion, and drug response, Pigment Cell Melanoma Res. {\bf 27}, 764 
(2014).

\item E. Mehrara, E. Forssell-Aronsson, A. Ahlman, and P. Bernhardt, Specific growth rate versus doubling time 
for quantitative characterization of tumor growth rate, Cancer Res. {\bf 67}, 3970 (2007).

\item P. Haridas, C.J. Penington, J.A. McGovern, D.S. McElwain, and M.J. Simpson, Quantifying rates of cell migration 
and cell proliferation in co-culture barrier assays reveals how skin and melanoma cells interact during melanoma 
spreading and invasion, J. Theor. Biol. {\bf 423}, 13 (2017).

\item A.A. Anderson, A hybrid mathematical model of solid tumour invasion: the importance of cell adhesion, Math. Med. 
Biol. {\bf 22}, 163 (2005).

\item E. Brunet and B. Derrida, Shift in the velocity of a front due to a cutoff, Phys. Rev. E {\bf 56}, 2597 (1997).

\item M. Scianna and L.A. Preziosi, Hybrid model describing different morphologies of tumor invasion fronts, Math. 
Model. Nat. Phenom. {\bf 7}, 78 (2012).

\item E. Khain, M. Katakowski, N. Charteris, F. Jiang, and M. Chopp, Migration of adhesive glioma cells: Front 
propagation and fingering, Phys. Rev. E {\bf 86}, 011904 (2012).

\item L.L. Munn, Dynamics of tissue topology during cancer invasion and metastasis, Phys. Biol. {\bf 10}, 065003 
(2013).

\item D.C. Markham, M.J. Simpson, P.K. Maini, E.A. Gaffney, and R.E. Baker, Comparing methods for modelling spreading 
cell fronts, J. Theor. Biol. {\bf 353}, 95 (2014).

\item N. Podewitz, F. J\"ulicher, G. Gompper, and J. Elgeti, Interface dynamics of competing tissues, New J. Phys. 
{\bf 18}, 083020 (2016).

\item B. Dey, G.R. Sekhar, and S.K. Mukhopadhyay, In-vivo mimicking model for solid tumor towards hydromechanics of 
tissue deformation and creation of necrosis, J. Biol. Phys. {\bf 44}, 361 (2018).

\item T.H. Keitt, Spectral representation of neutral landscapes, Landscape Ecol. {\bf 15}, 479 (2000).

\item M. Sahimi and S.E. Tajer, Self-affine fractal distributions of the bulk density, elastic moduli, and seismic 
wave velocities of rock, Phys. Rev. E {\bf 71}, 046301 (2005).

\item S.M. Vaez Allaei and M. Sahimi, Shape of a wave front in a heterogenous medium, Phys. Rev. Lett. {\bf 96}, 
075507 (2006).

\item F. Ghasemi, J. Peinke, M. Sahimi, and M.R. Rahimi Tabar, Regeneration of stochastic processes: An inverse 
method, Europ. Phys. J. B {\bf 47}, 411 (2005).

\item F. Ghasemi, M. Sahimi, J. Peinke, and M.R. Rahimi Tabar, Analysis of non-stationary data for heart-rate 
fluctuations in terms of drift and diffusion coefficients, J. Biol. Phys. {\bf 32}, 117 (2006).

\item B.B. Mandelbrot and J.W. van Ness, Fractional Brownian motion, fractional Guassian noise, and their applications,
SIAM Rev. {\bf 10}, 422 (1968).

\item M. Pr\"ahofer and H. Spohn, Universal distributions for growth processes in 1 + 1 dimensions and random 
matrices, Phys. Rev. Lett. {\bf 84}, 4882 (2000).

\item T. Sasamoto and H. Spohn, One-dimensional Kardar-Parisi-Zhang equation: An exact solution and its 
universality, Phys. Rev. Lett. {\bf 104}, 230602 (2010).

\item G. Amir, I. Corwin, and J. Quastel, Probability distribution of the free energy of the continuum directed random 
polymer in 1 + 1 dimensions, Commun. Pure Appl. Math. {\bf 64}, 466 (2011).

\item P. Calabrese and P. Le Doussal, Exact solution for the Kardar-Parisi-Zhang equation with flat initial conditions,
Phys. Rev. Lett. {\bf 106}, 250603 (2011).

\item A.A. Saberi, H. Dashti-Naserabadi, and J. Krug, Competing universalities in Kardar-Parisi-Zhang growth models,
Phys. Rev. Lett. {\bf 122}, 040605 (2019).

\end{list}%

\end{document}